\documentclass{PoS}
\usepackage{import}

\usepackage{amsmath}

\DeclareMathOperator\erf{erf}
\DeclareMathOperator\F{F}
\DeclareMathOperator\U{U}

\title{On the On-Off Problem:\\ An Objective Bayesian Analysis}

\ShortTitle{Objective Bayesian On-Off Analysis}

\author{\speaker{Max Ludwig Ahnen}\\
        ETH Zurich, Institute for Particle Physics, Otto-Stern-Weg 5, 8093 Zurich, Switzerland\\
        E-mail: \email{mahnen@phys.ethz.ch}}

\abstract{The On-Off problem, aka. Li-Ma problem, is a statistical problem where
a measured rate is the sum of two parts. The first is due to a signal and the
second due to a background, both of which are unknown. Mostly frequentist
solutions are being used that are only adequate for high count numbers.
When the events are rare such an approximation is not good enough. Indeed, in
high-energy astrophysics this is often the rule rather than the exception.

I will present a universal objective Bayesian solution that depends only on the
initial three parameters of the On-Off problem: the number of events in the
``on'' region, the number of events in the ``off'' region, and their
ratio-of-exposure.

With a two-step approach it is possible to infer the signal's significance,
strength, uncertainty or upper limit in a unified a way. The approach is valid
without restrictions for any count number including zero and may be widely
applied in particle physics, cosmic-ray physics and high-energy astrophysics. I
apply the method to Gamma Ray Burst data.}

\FullConference{The 34th International Cosmic Ray Conference,\\
		30 July- 6 August, 2015\\
		The Hague, The Netherlands}

\begin{document}

\section{Introduction}
\label{sec:introduction}
\begin{figure}[ht]
\centering
\includegraphics[width=1\textwidth]{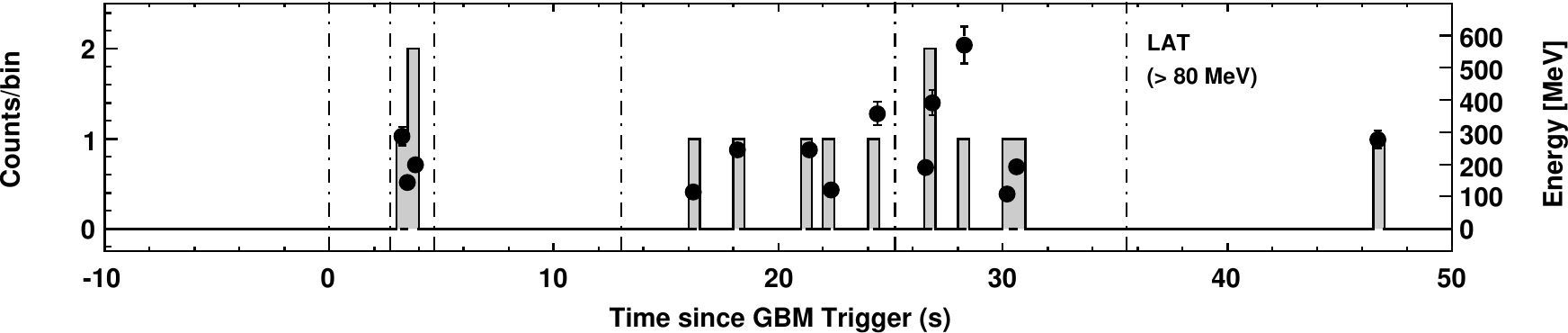}
\caption{\label{fig:latgrb}A typical low count high-energy astrophysics data set.
It shows gamma rays measured from GRB080825C as observed by Fermi-LAT. Black dots
represent the energy measurement, the gray bars represent the number of photons. Figure reproduced from \cite{abdo2009fermi}}
\end{figure}

Typical counting experiments measure discrete sets of events. 
Such data are often \cite{li1983analysis,cousins2008evaluation} modeled with the 
Poisson distribution. The Poisson distribution may be approximated by a normal distribution 
when measuring many events. However, when data is rare, such an approximation is not 
good enough. In Fig. \ref{fig:latgrb}, a typical example of a low count data sample is shown.
In this case it is an observation of a Gamma Ray Burst (GRB) with the Fermi-LAT instrument. 
The question arises: What do you do when it simply impossible to ``go out and get more data''?

\section{The On-Off Problem}
In the On-Off problem, also known as Li-Ma problem,
one would like to infer a signal rate in the presence of an imprecisely known background rate. 
The measurement consists of the observation of $N_{\mathrm{on}}$ events in some ``on'' region
with a potential signal and $N_{\mathrm{off}}$ events in some ``off'' region, known to be 
signal free.
Additionally to the number counts in the on-and off regions, there is a third parameter. This parameter
is the ratio $\alpha$ of exposures for the ``on'' and ``off'' regions and taken to be known with negligible uncertainty. 
In the case of gamma ray astronomy, Berge et al.~\cite{Berge2007} explain the problem and its parameters well.

The common frequentist analyses, based on likelihood ratios and
other methods \cite{li1983analysis,cousins2008evaluation,rolke2005limits}
often assume normal distributed random numbers and therefore
lose their foundation when applying them to low count numbers. 
They also get into trouble at the border of the physical parameter space \cite{rolke2005limits}.

The common Bayesian solutions to the On-Off problem 
are either subjective Bayesian (using proper posteriors \cite{gregory2005bayesian}) or they
avoid specifying the alternative hypothesis at all by using some tail-area 
probability inference in the spirit of p-values \cite{gillessen2005significance}.
Subjective Bayesian methods usually introduce a fourth, subjective, parameter (often the upper limit
to the signal parameter) which makes the probability statement somewhat dependent on the 
individual physicist. The tail-area methods, besides ignoring the beauty of Bayesian 
hypothesis testing, seem to overestimate the probability~\cite{gregory2005bayesian}.

Most of all none of the methods in the literature so far cover the full range of the problem:
First, the probability that the observed counts are due to 
background only has to be calculated. Second, 
the signal contribution has to be estimated.
In this proceeding I present a two-step objective Bayesian solution to the full On-Off problem~\cite{knoetig2014signal}, 
inspired by \cite{caldwell2006signal}, that addresses these 
issues in a unified a way. 

\section{Methods Development}
\label{sec:methods}
The idea behind objective Bayesian analysis simple. One takes, in a sense, ``flat'' priors
representing the lack of knowledge. These are usually improper (do 
not integrate to one). However, in combination with Bayes theorem 
these can be used to produce proper posteriors and answer a basic what-if question:
What is the result if the data were dominant? 

One particularly popular 
objective Bayesian prior is Jeffreys's rule \cite{jeffreys1961theory,pdg}. 
Jeffreys was motivated by invariance requirements and suggested to take 
a specific objective prior to make the result (the posterior) invariant 
under re-parametrization. This prior is a keystone in this analysis. 

The analysis follows the method outlined by \cite
{caldwell2006signal} and is done in two steps.
First, the odds that the observed counts are due to the background model
are calculated. If this is smaller than a previously 
defined value, the signal is said to be detected. Second, 
the signal contribution or upper limit is calculated, 
depending on whether the detection limit has been reached.
The first is done with objective Bayesian hypothesis testing via Bayes factors. 
The second is done with objective Bayesian estimation.

\subsection{First step: Hypothesis testing}
One problem that appears is that objective priors are only
defined up to a proportionality constant and those constants become relevant in this case. 
A full discussion of the topic can be found in \cite{knoetig2014signal,ahnen2014forum}. In short, 
there is no generally agreed objective Bayesian hypothesis testing.
I suggest to use a method sometimes called ``minimal sample device'' \cite{spiegelhalter1981,ghosh2002},
in order to fix the issue with the proportionality constants. 
The evaluation of these assumptions in the case of the On-Off problem
can be found in \cite{knoetig2014signal,ahnen2014forum}.
After all, the odds of the background model over the signal 
model are
 \begin{eqnarray}
 B_{\mathrm{01}} &=& \frac{c_{\mathrm{0}}}{c_{\mathrm{1}}} \frac{\gamma}{\delta} ,
 \label{eqn:bayesfactor}
 \end{eqnarray}
where $\gamma$ and $\delta$ are defined using the Gamma function
$\Gamma\left(x\right)$ and the hypergeometric function
$_{2}\F_{1}\left(a,b;c;z\right)$:
\begin{eqnarray}
\gamma & := & \left(1+2N_{\mathrm{off}}\right) \alpha^{\frac
{1}{2}+N_{\mathrm{on}}+N_{\mathrm{off}}} \Gamma\left(\frac{1}{2}+N_{\mathrm{on}}+N_{\mathrm{off}}
\right) \nonumber\\ \delta & := & 2\left(1+\alpha\right)^{N_{\mathrm
{on}}+N_{\mathrm{off}}} \Gamma\left(1+N_{\mathrm{on}}+N_{\mathrm{off}} 
\right) _{2}\F_{1}\left(\frac{1}{2}+N_{\mathrm
{off}},1+N_{\mathrm{on}}+N_{\mathrm{off}};\frac{3}{2}+N_{\mathrm
{off}} ;-\frac{1}{\alpha}\right) \nonumber \\
\frac{c_0}{c_1} & = & \frac{2 \arctan\left(\frac{1}{\sqrt{\alpha}}\right)}{\sqrt{\pi}} .
 \end{eqnarray}
A signal detection based on Eqn. \ref{eqn:bayesfactor} may be claimed when the resulting odds 
of the background model are low. 
I propose to use use a ''Bayesian z-value``, similar to~\cite{gillessen2005significance}
 \begin{equation}
S_{\mathrm{b}}=\sqrt{2}  \erf^{-1}\left(1-B_{\mathrm{01}}\right) ,
\label{eqn:sigb}
\end{equation}
where $B_{\mathrm{01}} = 5.7\times10^{-7}$ would correspond to $S_{\mathrm{b}}=5$ or ''5 sigma``. 
This definition allows for an easy comparison with frequentist significance methods~
\cite{knoetig2014signal,ahnen2014forum}. However one must keep in mind that 
the odds of a model and the frequency of
an outcome are two different things. $B_{\mathrm{01}}$ 
explicitly weighs alternative models, while the
frequentist methods do not.

\subsection{Second Step: Signal Estimation}
After determining the Bayes factor of the background model over the signal model, one proceeds to 
estimating the signal contribution. This is done via objective Bayesian estimation.
If the data show a significant detection the signal model can 
be assumed to be true. Then, the most probable signal parameter value should be calculated 
and a physical error interval should be given. 
If the data show no significant detection, an upper limit on the 
signal parameter should be calculated, assuming that the signal is 
there (i.e. the signal model is true) but too weak to be measured.
In both cases one needs the conditional probability $P\left(\lambda_{\text{s}}|N_{\text{on}},N_{\text{off}},H_{
 \text{1}}\right)$ of the signal $\lambda_{\mathrm{s}}$, given the number counts and 
the signal model $H_{
 \text{1}}$. The improper prior is acceptable in this case 
because the proportionality constant $c_1$ cancels and the posterior is proper.
After marginalization over the background parameter $\lambda_{\text{bg}}$, the result 
is (calculation in \cite{knoetig2014signal})
 \begin{eqnarray}
 & & P\left(\lambda_{\text{s}}|N_{\text{on}},N_{\text{off}},H_{
 \text{1}}\right) = P_{\text{P}}\left(N_{\text{on}}+N_{\text
 {off}}|\lambda_{\text{s}}\right)  \label{eqn:posterior}
 \frac{\U\left[\frac{1}{2}+N_{\text{off}},1+N_{\text
 {off}}+N_{\text{on}},\left(1+\frac{1}{ \alpha}
 \right)\lambda_{\text{s}}\right]}{_{2}\tilde{\F}_{1}
 \left(\frac{1}{2}+N_{\text{off}} ,1+N_{\text{off}}+N_{\text
 {on}};\frac{3}{2}+N_{\text{off}};-\frac{1}{\alpha}
 \right)} ,  
\end{eqnarray}
as expressed in terms of three functions, 
namely the Poisson distribution 
$P_{\text{P}}\left(N|\lambda\right)$, the regularized hypergeometric 
function $_{2}\tilde{\F}_{1}\left(a,b;c;z\right) = 
\frac{_{2}\F_{1}\left(a,b;c;z\right)}{\Gamma\left(c\right)}$, and 
the Tricomi confluent hypergeometric function $\U\left(a,b,z\right)$.
This posterior contains the full signal parameter information. 
In order to state a flux, one should take the mode $\lambda_{\text{s}}^*$,
of the posterior distribution 
$P\left(\lambda_{\text{s}}|N_{\text{on}},N_{\text{off}},H_{\text{1}}
\right)$, as signal estimator. The error on the signal estimator can be 
evaluated numerically from the cumulative distribution function. On interesting choice is 
the highest posterior density interval (HPD) $[\lambda_{\text{min}},\lambda_{\text{max}}]$~\cite{knoetig2014signal}
containing 68\% probability, calculated as
\begin{equation}
 \int_{\lambda_{\text{min}}}^{\lambda_{\text{max}}}P
 \left(\lambda_{\text{s}}|N_{\text{on}},N_{\text{off}},H_{\text
 {1}}\right)d\lambda_{\text{s}} = 0.68 ,
\end{equation}
together with the constraint
\begin{equation}
 P\left(\lambda_{\text{min}}|N_{\text{on}},N_{\text{off}},H_{
 \text{1}}\right) = P\left(\lambda_{\text{max}}|N_{\text{on}},N_{
 \text{off}},H_{\text{1}}\right) .
\end{equation}

In case an upper limit should be calculated one can solve the cumulative 
distribution function, for instance, for a $99\%$ probability limit $\lambda_{99}$ on the signal 
parameter $\lambda_{\text{s}}$ as
\begin{equation}
 \int_{0}^{\lambda_{99}}P\left(\lambda_{\text{s}}|N_{\text
 {on}},N_{\text{off}},H_{\text{1}}\right)d\lambda_{\text{s}} 
 =0.99 . \label{eqn:l99}
\end{equation}
These results are natural in a Bayesian approach of the problem 
but hard to calculate in a frequentist approach. 
Most frequentist methods particularly struggle with the
marginalization and fail at the border of the 
parameter space. In this approach, all possible number counts are 
dealt with in a uniform way, no matter if zero counts or thousands
of counts. A further benefit is that the signal posterior probability intervals are always
physically meaningful
(i.e. positive
$\lambda_{\text{s}}^{*},\lambda_{min},\lambda_{max},\lambda_{99},...$).

\label{sec:application}
As an application of the developed method, I want to demonstrate the calculations on two 
measurements of gamma rays from GRBs. These are extraterrestrial flashes of gamma rays 
mostly lasting only a few seconds. One interesting question is how exactly GRBs produce high energy gamma rays
~\cite{abdo2009fermi}. Because of the GRBs duration and fluence, satellites and Cherenkov telescopes measure 
only few events during the flare itself or shortly after. 

\begin{figure}[ht]
\centering
\includegraphics[width=0.6\textwidth]{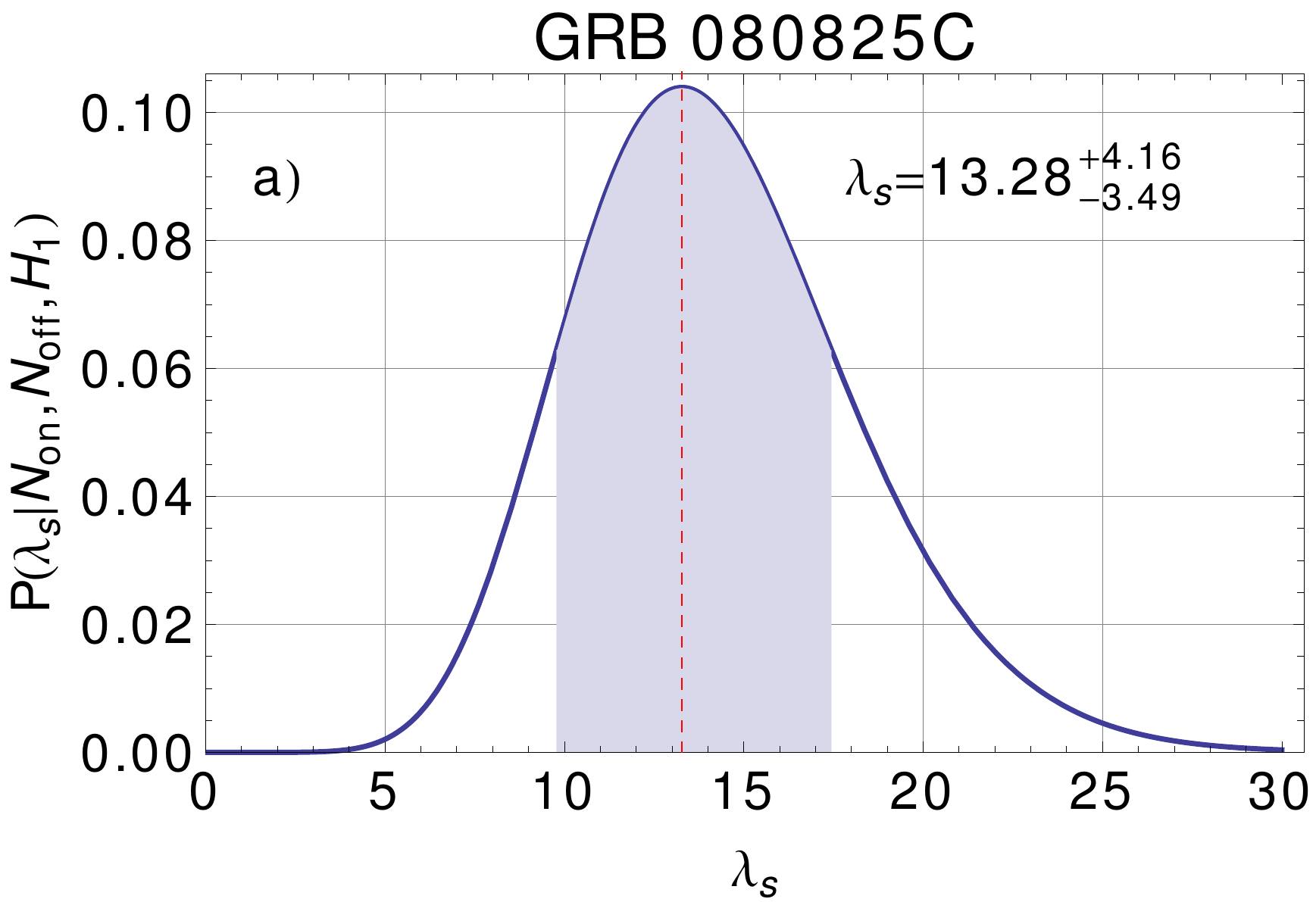}
\caption{\label{fig:signal}The conditional probability $P\left(\lambda_{\text{s}}|N_{\text{on}},N_{\text{off}},H_{
 \text{1}}\right)$ of the signal $\lambda_{\mathrm{s}}$, given the Fermi-LAT number counts
 of GRB080825C. The blue band indicates the HPD interval for the signal parameter posterior probability. 
 Figure reproduced from \cite{knoetig2014signal}.}
\end{figure}

The first example is the GRB080825C as seen
by Fermi-LAT (see Sec.~\ref{sec:introduction}). In Fig.~\ref{fig:latgrb} 
one sees the light curve in gamma rays > 80MeV. In total there were $N_{\text{on}} = 15$ 
on events and $N_{\text{off}} = 19$ off events, with an exposure ratio of $\alpha = 33/525$~\cite{abdo2009fermi}.
Running the numbers shows that
the odds of the background model over the signal model are (Eqn.~\ref{eqn:bayesfactor}) 
$B_{01}=9.66\times10^{-10}$, or as expressed with the nonlinear 
scale of Eqn.~\ref{eqn:sigb}, $S_{\text{b}}= 6.11$. These numbers compare well to 
the published value of \mbox{$S_{\text{li-ma}}=6.4$} \cite{abdo2009fermi}. 
Clearly, the odds of the background model are low and the
GRB is therefore detected. Then, in the second step, one performs the signal estimation. 
The result is plotted in Fig.~\ref{fig:signal}.
\begin{figure}[ht]
\centering
\includegraphics[width=0.6\textwidth]{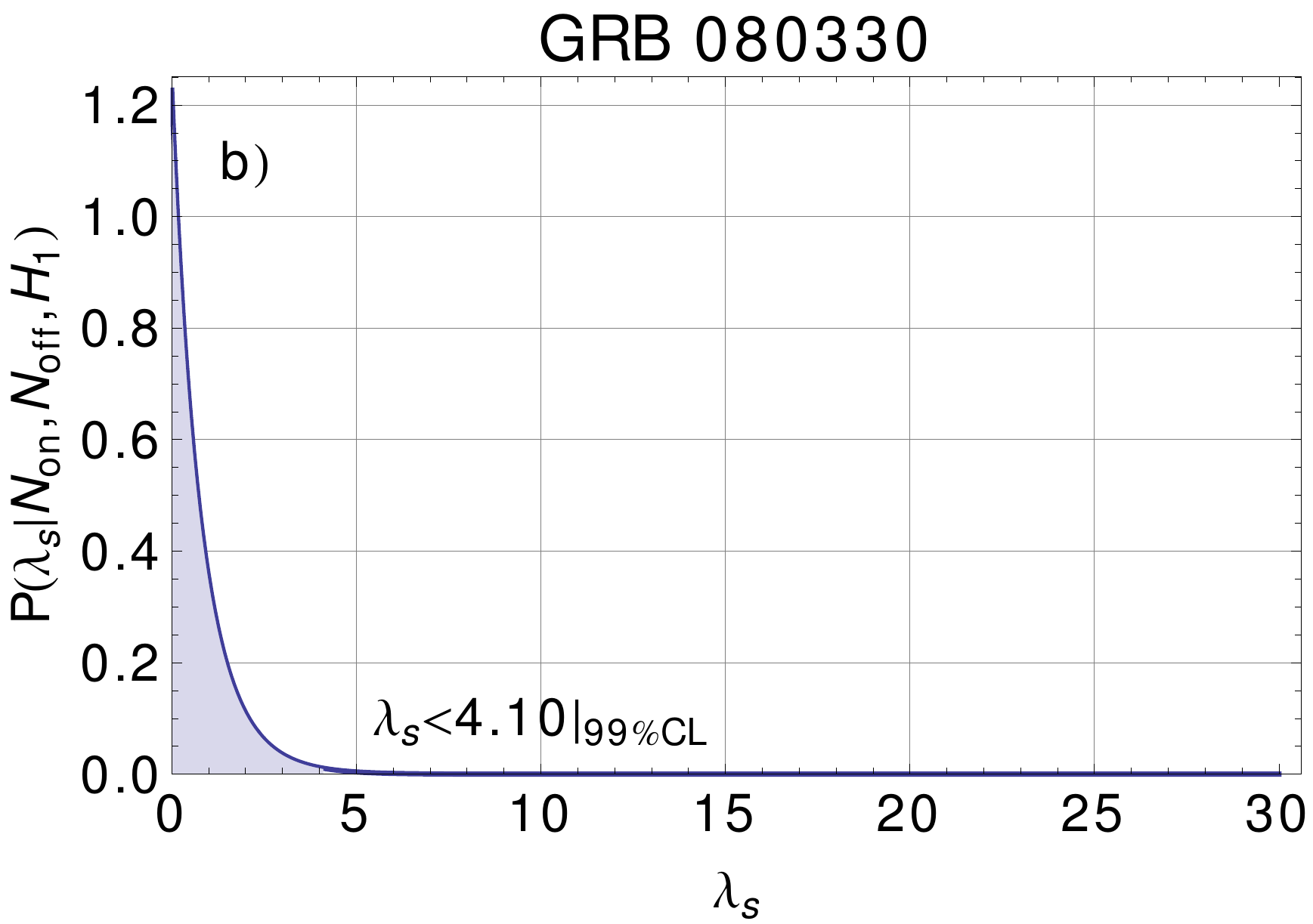}
\caption{\label{fig:ul}The conditional probability $P\left(\lambda_{\text{s}}|N_{\text{on}},N_{\text{off}},H_{
 \text{1}}\right)$ of the signal $\lambda_{\mathrm{s}}$, given the VERITAS number counts
 of GRB080330. The blue band indicates the 99\% probability upper limit to the 
 signal parameter. Figure reproduced from \cite{knoetig2014signal}.}
\end{figure}
The result of $\lambda_{\text{s}} = 13.28_{-3.49}^{+4.16}$ is in good agreement
with the published reference of $\lambda_{\text{Ref.}}=13.7$. 

The second example is the GRB080330 as observed
by the VERITAS Cherenkov telescope~\cite{acciari2011veritas}. 
The measurement shows $N_{\text{on}} = 0$ 
on events and $N_{\text{off}} = 15$ off events, with an exposure ratio of $\alpha = 0.123$~\cite{knoetig2014signal}.
The corresponding odds of the background model are $B_{01}=2.29$, unsurprisingly favoring the 
null hypothesis as not a single on event was detected. 
Now, assuming that the source is there one can put an upper limit to $\lambda_{\text{s}}$. The result
is plotted in Fig.~\ref{fig:ul}. 
The published value uses a frequentist upper limit setting method, popularized by Rolke et al.~\cite{rolke2005limits}.
Their result is $\lambda_{99}^{\textrm{Rolke}} = 2.4$~\cite{knoetig2014signal}, which somewhat lower than the number from Eqn.~\ref{eqn:l99}, 
$\lambda_{99} = 4.10$.
A detailed analysis indicates~\cite{knoetig2014signal} that, especially at the border of the parameter space 
for $N_{\text{on}} \leq \alpha N_{\text{off}}$, Rolke's method is an overestimation and therefore limited. 
These limits are overcome by the Bayesian method.

\section{Validation and Discussion}
\label{sec:discussion}
In order to validate the method and to check if the assumptions made are 
sensible, an extensive validation was made~\cite{knoetig2014signal}. 
The validation shows that the two-step method behaves well in all test-case examples, 
in particular at $N_{\textrm{on}} \sim \alpha N_{\textrm{off}}$.
The objective Bayesian hypothesis testing converges to the results from other methods for high count numbers.
The objective Bayesian signal estimation can reconstruct the true signal parameter $\lambda_{\textrm{s}}$
with a good error estimate.

\section{Conclusion}
\label{sec:conclusion}
Claiming detections, setting credibility intervals, or setting upper limits can be unified 
over the whole On-Off problem parameter range in one consistent two-step objective Bayesian method.
An example implementation in Python can be downloaded from the public git-repository \\
\href{https://bitbucket.org/mknoetig/obayes_onoff_problem}{https://bitbucket.org/mknoetig/obayes\_onoff\_problem}.

\bibliographystyle{JHEP}
\bibliography{mahnen_proceeding}

\providecommand{\href}[2]{#2}\begingroup\raggedright\begin{thebibliography}{10}

\bibitem{li1983analysis}
T.~P. Li and Y.~Q. Ma, {\it Analysis methods for results in gamma-ray
  astronomy},  {\em Astrophys. J.} {\bf 272} (1983) 317--324.

\bibitem{cousins2008evaluation}
R.~D. Cousins, J.~T. Linnemann, and J.~Tucker, {\it Evaluation of three methods
  for calculating statistical significance when incorporating a systematic
  uncertainty into a test of the background-only hypothesis for a poisson
  process},  {\em Nucl. Instrum. Methods A} {\bf 595} (2008), no.~2 480--501.

\bibitem{abdo2009fermi}
{\bf Fermi LAT/GBM Collaborations} Collaboration, A.~Abdo et~al., {\it Fermi
  observations of high-energy gamma-ray emission from grb 080825c},  {\em
  Astrophys. J.} {\bf 707} (2009), no.~1 580.

\bibitem{Berge2007}
D.~{Berge}, S.~{Funk}, and J.~{Hinton}, {\it {Background modelling in
  very-high-energy {$\gamma$}-ray astronomy}},  {\em Astron. Astrophys.} {\bf
  466} (May, 2007) 1219--1229.

\bibitem{rolke2005limits}
W.~A. Rolke, A.~M. L{\'o}pez, and J.~Conrad, {\it Limits and confidence
  intervals in the presence of nuisance parameters},  {\em Nucl. Instrum.
  Methods A} {\bf 551} (2005), no.~2 493--503.

\bibitem{gregory2005bayesian}
P.~Gregory, {\em Bayesian logical data analysis for the physical sciences}.
\newblock Cambridge University Press, 2005.

\bibitem{gillessen2005significance}
S.~{Gillessen} and H.~L. {Harney}, {\it {Significance in gamma-ray astronomy -
  the Li - Ma problem in Bayesian statistics}},  {\em Astron. Astrophys.} {\bf
  430} (2005) 355--362.

\bibitem{knoetig2014signal}
M.~L. Knoetig, {\it Signal discovery, limits, and uncertainties with sparse
  on/off measurements: an objective bayesian analysis},  {\em Astrophys. J.}
  {\bf 790} (2014), no.~2 106.

\bibitem{caldwell2006signal}
A.~Caldwell and K.~Kr{\"o}ninger, {\it Signal discovery in sparse spectra: A
  bayesian analysis},  {\em Phys. Rev. D} {\bf 74} (2006), no.~9 092003.

\bibitem{jeffreys1961theory}
H.~Jeffreys, {\em Theory of probability}.
\newblock Clarendon Press Oxford, 1961.

\bibitem{pdg}
{\bf Particle Data Group} Collaboration, J.~Beringer et~al., {\it Review of
  particle physics},  {\em Phys. Rev. D} {\bf 86} (2012) 010001.

\bibitem{ahnen2014forum}
M.~L. Ahnen, {\it On the on/off problem},  in {\em Bayes Forum Munich}, 2014.

\bibitem{spiegelhalter1981}
D.~J. Spiegelhalter and A.~F.~M. Smith, {\it Bayes factors for linear and
  log-linear models with vague prior information},  {\em J. R. Statist. Soc. B}
  {\bf 44} (1981), no.~3 377--387.

\bibitem{ghosh2002}
J.~K. Ghosh and T.~Samanta, {\it Nonsubjective bayes testing --- an overview},
  {\em J. Statist. Plann. Inference} {\bf 103} (2002) 205--223.

\bibitem{acciari2011veritas}
{\bf VERITAS Collaboration} Collaboration, V.~A. {Acciari} et~al., {\it
  {VERITAS Observations of Gamma-Ray Bursts Detected by Swift}},  {\em
  Astrophys. J.} {\bf 743} (2011) 62.

\end{thebibliography}\endgroup

\end{document}